\documentclass[preprint,11pt]{revtex4-1}

\usepackage[utf8x]{inputenc}
\usepackage[english]{babel}

\usepackage[usenames,dvipsnames,svgnames]{xcolor}
\usepackage{amsmath}
\usepackage{amssymb}
\usepackage{graphicx}
\usepackage{times}
\usepackage{sidecap}

\newcommand{\KP}{Kronig-Penney}
\newcommand{\GP}{Gross-Pitaevskii}
\newcommand{\KPacr}{KP}

\newcommand{\GPEacr}{GPE}

\newcommand{\qmop}[1]{\hat{#1}}

\newcommand{\gpeham}{\hat{H}_{\mathrm{GP}}}

\newcommand{\KPperiod}{l}
\newcommand{\vzero}{V_0}

\newcommand{\kopt}{k_{\mathrm{OL}}}
\newcommand{\lambdaopt}{\lambda_{\mathrm{OL}}}
\newcommand{\Er}{E_{\mathrm{R}}}
\newcommand{\VKP}{V_{\mathrm{KP}}}
\newcommand{\chempot}{\mu}

\newcommand{\gpwf}{{ \Phi }}

\newcommand{\gpwfk}{\phi_{k}}
\newcommand{\gpwfkw}{\gpwfk^{\tiny\mathrm{w}}}
\newcommand{\gpwfkb}{\gpwfk^{\tiny\mathrm{b}}}

\newcommand{\gd}{{g}}
\newcommand{\density}[1][1]{n_{#1}}
\newcommand{\densityw}{\density^{\tiny\mathrm{w}}}
\newcommand{\densityb}{\density^{\tiny\mathrm{b}}}

\newcommand{\intfactor}{\gd \avgdensity}
\newcommand{\gpenergy}[1][]{E_{#1}}

\newcommand{\phase}{{S}}
\newcommand{\phasew}{S^{\tiny\mathrm{w}}}
\newcommand{\phaseb}{S^{\tiny\mathrm{b}}}

\newcommand{\avgdensity}{{n}}

\newcommand{\jacobisn}{\mathrm{sn}}

\newcommand{\ellipm}{m_\mathrm{j}}

\newcommand{\densitydl}[1][1]{\tilde{n}_{#1}}

\begin{document}

\title{Periodic Ultranarrow Rods as 1D Subwavelength Optical Lattices}

\author{Omar Abel Rodríguez-López}
\affiliation{Institute of Physics, UNAM, 04510 México City, México}
\email{oarodriguez.mx@gmail.com}
% \homepage[ORCID: ]{https://orcid.org/0000-0002-3635-9248}

\author{M. A. Solís.}
\affiliation{Institute of Physics, UNAM, Apdo. Postal 20--364, 01000 México}
\email{masolis@fisica.unam.mx}

\date{Modified: \today / Compiled: \today}

\begin{abstract}
  We report on ground state properties of a one-dimensional, weakly interacting
  Bose gas constrained by an infinite multi-rods periodic structure at zero
  temperature. We solve the stationary Gross-Pitaevskii equation (GPE) to obtain
  the Bloch wave functions from which we give a semi-analytical solution for the
  density profile, as well as for the phase of the wave function in terms of the
  Jacobi elliptic functions, and the incomplete elliptic integrals of the first,
  second and third kind. Then, we determine numerically the energy of the ground
  state, the chemical potential and the compressibility of the condensate and
  show their dependence on the potential height, as well as on the interaction
  between the bosons. We show the appearance of loops in the energy band
  spectrum of the system for strong enough interactions, which appear at the
  edges of the first Brillouin zone for odd bands and at the center for even
  bands. We apply our model to predict the energy band structure of the Bose gas
  in an optical lattice with subwavelength spatial structure. To discuss the
  density range of the validity of the GPE predictions, we calculate the ground
  state energies of the free Bose gas using the GPE, which we compare with the
  Lieb-Liniger exact energies.


\end{abstract}

\maketitle

\section{Introduction}%
\label{sec:introduction}

Since the realization of a Bose-Einstein condensate (BEC) in
1995~\cite{bib:anderson_science.1995,bib:davis_prl.1995}, many new types of
experiments have been proposed and realized involving BEC.\ In particular,  we
know that a successful research line has been to study BEC within periodic
optical lattices created by superposition of two opposing laser lights in one,
two or three mutually perpendicular
directions~\cite{bib:jaksch_prl-81.3108.1998,bib:bloch_nat-phys-1.23.2005}, such
that the atomic gases are trapped in 3D multilayers, multitubes, or in a simple
cubic array of dots, respectively. Atoms are trapped in one direction by the
standing wave formed by two opposite laser lights whose effective potential
acting on atoms has the generic sinusoidal form $V(x) = A  \sin^2( 2\pi \,
  x/\lambda)$, where $\lambda $ is the wavelength of the laser light and $A$ is
the lattice potential height given in energy units. This ability to generate
optical lattices of various types has become a fundamental tool to study the
physics of bosonic or/and fermionic many atoms
systems~\cite{bib:gross_science.357.995.2017}. For instance, the superfluid-Mott
insulator quantum phase transitions were experimentally probed ``in a
Bose–Einstein condensate with repulsive interactions, held in a
three-dimensional optical lattice
potential''~\cite{bib:jaksch_prl-81.3108.1998,bib:greiner_nature.2002}, as well
as in a 1D optical lattice~\cite{bib:haller_nature.2010}. However, these kind of
optical lattices are limited by its spatial resolution, which is of the order
$\lambda$, to manipulate atoms. Fortunately, there has recently been a notorious
interest and advances in developing tools to overcome the diffraction limit,
arriving to the physical realization of subwavelength optical lattices (SWOL) of
\emph{nearly} $\delta$-function potential with ultranarrow barriers of width
below
$\lambda/50$~\cite{bib:lacki_prl-117.233001.2016,bib:wang_prl-120.083601.2018};
reported results include, among others, the energy band structure. These SWOL
can be seen as a very close experimental realization of the Dirac comb (DC)
potential~\cite{bib:merzbacher_qm.jws.2nd.1969}, as well as attractive setups to
achieve $p$-wave superfluidity in a gas of fermions in 2D optical
lattices~\cite{bib:fedorov_pra.2017}.

From the theoretical point of view, the effect of a point-like, Dirac delta
potential on the weakly-interacting Bose gas has been
studied~\cite{bib:seaman-pra-3.033609.2005}. Also, there has been substantial
research on the properties of the Bose gas within a Dirac comb potential in the
mean-field approximation, for instance, on the behavior of density profile of
the condensate (both analytically and numerically), and how the interactions can
have a profound impact on the energy spectrum, like the appearance of ``swallow
tails'' in the band
structure~\cite{bib:theodorakis_jpa-13.4835.1997,bib:machholm_pra-67.053613.2003,bib:weidong-li_phys-rev-e-70.016605.2004,bib:seaman-pra-3.033622.2005},
as well as analysis on the influence of the periodic structure on the stability
of superfluid currents~\cite{bib:dong_lp-2.190.2007,bib:danshita_pra.2007}.
Also, the DC potential has been extended to model interactions between atoms and
ions forming a lattice~\cite{bib:negretti_prb.2014} where both $s$-wave and
$p$-wave scattering is present. However, to our knowledge, research on the
properties of the Bose gas within the true Kronig-Penney (KP)
potential~\cite{bib:kronig-prsa-130.499.1931} with barriers and wells is
missing, since all of the referred previous works employ the KP potential in the
limit when barriers become \emph{exact} Dirac deltas, i.e., the so-called DC
potential.

In this work we study a one-dimensional weakly-interacting Bose gas within an
infinite permeable multi-rods periodic structure which we use to discuss the
ultranarrow rods limit as a 1D optical lattice with subwavelength spatial
structure. The structure is modeled by a KP potential, which we analytically
solve in the weak interaction regime where the
GPE~\cite{bib:gross_inc.1961,bib:pitaevskii_spjetp.1961} is applicable. The KP
potential, i.e., periodic structure of well plus barrier, has the advantage that
it is closer to the sinusoidal optical lattice potential than the Dirac comb,
but at the same time it retains the simplicity to be solved analytically. We
analyze the effects of the height and width of the barriers, as well as the
interaction strength between bosons, on the ground state properties such as the
density profile, the chemical potential, and the energy spectrum.
Then we use our model to represent SWOL by means of very high, narrow barriers
which retain their finite extent. In other words, although the barriers are very
narrow compared to the optical lattice period, they have a nonzero width. The
finite extent of the barriers holds even for future prospects of SWOL with
period $\lambda/4$~\cite{bib:subhankar_arxiv.2019}.

On the other hand, the Bose gas within a multirods periodic structure is the
same as the beautiful exact soluble Lieb-Liniger (LL)
model~\cite{bib:lieb_phys-rev-130.1605.1963} but within an external KP
potential, which we analytically solve in the weakly interacting regime where the
Gross-Pitaevskii equation is applicable. We show the effects of the KP potential
on the ground state energy of the LL Bose gas, recovering the LL results in the
weak interaction regime when we delete the KP potential.  In order to establish
the density regime of the GPE applicability, given an interaction magnitude, we
use both ground state energies of the free Bose gas calculated exactly (LL case)
and approximately (GP case), to fix the density regime where both energies are
approximately equal. Correspondingly, for the case of the trapped Bose gas we give a
lower estimate for the average linear density of $250$ bosons per potential
period within the interaction interval used.

This work is developed in the following way. In
Sec.~\ref{sec:bose-gas-in-multi-rods} we present our model of a 1D Bose gas
within permeable multi-rods; we establish the boundary conditions to obtain the
constants on which the solutions depend. In Sec.~\ref{sec:density-profile} we
give the ground state density profiles and we calculate the chemical potential
and compressibility. In Sec.~\ref{sec:energy-spectrum} we calculate and plot the
nonlinear energy band structure where the most remarkable thing is the
appearance of energy loops, also known as swallow tails. In
Sec.~\ref{sec:sub-wavelength-opt-latt-kp} we show the behavior of the density
profile and the energy spectrum in the limit of very narrow barriers. We employ
our model to predict the energy spectrum of a interacting Bose gas within an
optical lattice with subwavelength spatial structure. Finally in the
Sec.~\ref{sec:conclusions} we give our conclusions.

\section{Bose gas within permeable multi-rods}%
\label{sec:bose-gas-in-multi-rods}

We study a one-dimensional, weakly-interacting Bose gas constrained by a
periodical structure composed of an infinite sequence of permeable rods of
length $b$, separated a distance $a$; the rods repeat along the $z$ direction.
We consider that the interactions between bosons are weak enough so that the
physical properties of our system can be correctly described by the
GPE~\cite{bib:pitaevskii.oup.2016}
\begin{equation}
  \label{eq:gpe-def}
  i \hbar \, \partial_t \Psi(z, t) = \gpeham \Psi(z, t),
\end{equation}
where $\Psi(z, t)$ is the wave function of the condensate. Since we are
interested in the stationary states, i.e., those that evolve in time like
\begin{equation}
  \label{eq:gpe-stat-state-def}
  \Psi(z, t) = \gpwf(z) e^{-i \mu t / \hbar},
\end{equation}
where $\mu$ is the chemical potential of the system, Eq.~\eqref{eq:gpe-def}
becomes the stationary Gross-Pitaevskii equation,
\begin{equation}
  \label{eq:stat-gpe-def}
  \gpeham \gpwf(z) = \mu \, \gpwf(z),
\end{equation}
where the l.h.s.~operator $\gpeham$ is the Gross-Pitaevskii time independent
hamiltonian
\begin{equation}
  \label{eq:gpe-hamiltonian}
  \gpeham = -\frac{\hbar^2}{2m}\partial_z^2 + V(z) + \gd |\gpwf(z)|^2,
\end{equation}
$m$ is the mass of the bosons, $\gd$ is the parameter that measures the strength
of the interaction between particles, and $V(z)$ is the external potential. The
stationary wave function $\gpwf(z)$ is subject to the normalization condition
\begin{equation}
  \label{eq:gpwf-norm}
  \int |\gpwf(z)|^2 \, dz = \mathcal{N},
\end{equation}
where $\mathcal{N}$ is the number of bosons in the condensate.

\begin{SCfigure}[50][b]
  \includegraphics[width=0.6\linewidth]{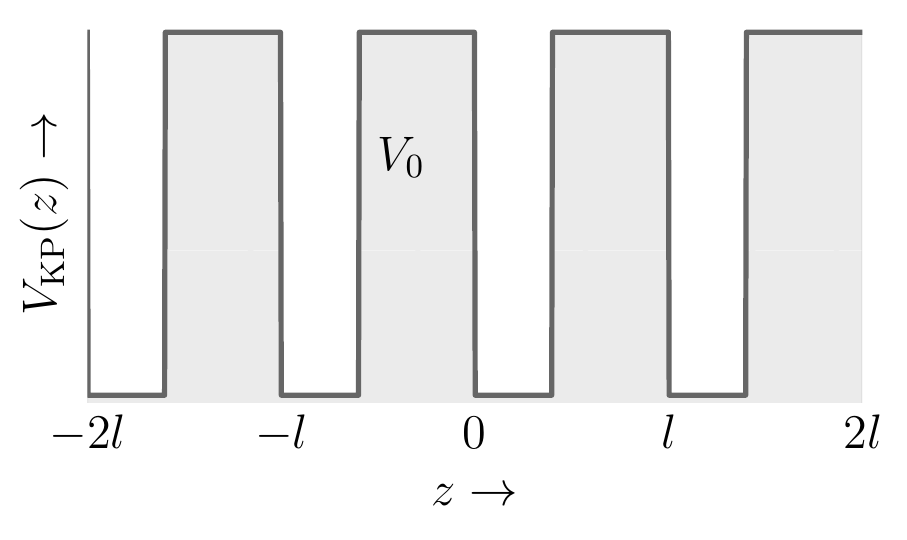}
  \caption{(Color online) Schema of the {\KP} potential $\VKP(z)$.}%
  \label{fig:kp-pot-schema}
\end{SCfigure}

The multi-rods structure is generated via an external Kronig-Penney  ({\KPacr})
potential~\cite{bib:kronig-prsa-130.499.1931} $V(z) \equiv \VKP(z)$. This
periodic potential is an array of barriers of width $b$ separated by a distance
$a$, each one with height $\vzero$, see Fig.~\ref{fig:kp-pot-schema}. The
  {\KPacr} potential can be written as
\begin{equation}
  \label{eq:kp-potential-def}
  \VKP(z) = \vzero \sum_{j=-\infty}^{\infty} \Theta[z -
    (j-1)l - a] \, \Theta[j l - z],
\end{equation}
where $\Theta(z)$ is the Heaviside step function and $l \equiv a + b$ the
potential period. For an infinite system like this, which repeats over and over,
the normalization condition can be defined within a single period $\KPperiod$,
in the following way
\begin{equation}
  \label{eq:gpwf-norm-period}
  \int_{0}^{l} |\gpwf(z)|^2 \, dz = N,
\end{equation}
where $N$ is the average number of bosons in the condensate over a length equal
to the potential period, such that the average linear density of the system
becomes $\avgdensity = N/l$. This condition fixes the value of the chemical
potential of the system, since the average number of bosons remains constant.
The energy per particle of the condensate can be defined in a similar way by
\begin{equation}
  \label{eq:gpe-energy}
  \frac{\gpenergy[][\gpwf]}{N}  =
  \frac{1}{N} \int_{0}^{l}                 \gpwf^{*}(z) \left[
    -\frac{\hbar^2}{2m}\partial_z^2 + \VKP(z) + \frac{\gd}{2} |\gpwf(z)|^2
    \right]  \gpwf(z) \, dz.
\end{equation}

Given the geometry of multi-rods, we can identify the potential period
$\KPperiod$ as a characteristic length which is either the distance between the
midpoints of any two consecutive barriers, or the distance between the midpoints
of two consecutive wells. This period $l$ is equal to that of an optical lattice
produced by two counter-propagating lasers with wavelength $\lambdaopt$ and wave
number $\kopt = 2\pi / \lambdaopt$, which is represented by an external
potential $V_\mathrm{OL}(z) = s \Er \sin^2(\kopt z)$, with $s$ being the lattice
height in recoil energy $E_R$ units, where $\Er \equiv \hbar^2 \kopt^2 / 2m$.
Since the period of the optical potential, i.e., the distance between two
consecutive maximums, is $l_\mathrm{OL} = \lambdaopt / 2$, and hence $\kopt =
  \pi / l_\mathrm{OL}$, the recoil energy of the optical lattice becomes $\Er =
  \hbar^2 \pi^2 / 2 m l_\mathrm{OL}^2$~\cite{bib:bloch_nat-phys-1.23.2005}. Doing
an analogy with the optical lattice, we can identify the recoil energy of our
multi-rods system as $\Er \equiv \hbar^2 \pi^2 / 2 m \KPperiod^2$, that
corresponds to the recoil energy of an optical lattice with the same period of
the {\KP} potential.

Since the density of our multi-rods system $|\gpwf(z)|^2$ is a physically
periodic function with the period of the {\KPacr} potential, $\gpeham =
(-\hbar^2 / 2m) \partial^2_z + \VKP(z) + g |\gpwf(z)|^2$ should be invariant
under translations by a distance $\KPperiod$, i.e., $\qmop{D}_\KPperiod \gpeham
\gpwf(z) = \gpeham \qmop{D}_\KPperiod \gpwf(z) = \chempot \qmop{D}_\KPperiod
\gpwf(z)$, where $\qmop{D}_\KPperiod$ is the translation operator whose action
is $\qmop{D}_\KPperiod f(z) = f(z + \KPperiod)$ for any function $f(z)$.
Therefore, the eigenstates of $\qmop{D}_\KPperiod$ are solutions of the
{\GPEacr} with chemical potential $\chempot$, i.e., they can be written as Bloch
waves:
\begin{equation}
  \label{eq:bloch-state}
  \gpwf_k(z) = e^{i k z} \, \gpwfk(z),
\end{equation}
where the function $\gpwfk(z)$ has the same periodicity as the potential, i.e.,
$\gpwfk(z + l) = \gpwfk(z)$, and $\hbar k$ is the quasi-momentum of bosons in
the condensate. Introducing~\eqref{eq:bloch-state} in~\eqref{eq:stat-gpe-def} we
obtain
\begin{equation}
  \label{eq:gpe-k}
  \gpeham^{k} \gpwfk(z) = \mu_k \gpwfk(z)
\end{equation}
with
\begin{equation}
  \label{eq:gpe-hamiltonian-k}
  \gpeham^{k} = \frac{\hbar^2}{2m} {\left(-i \partial_z + k \right)}^2 +
  \VKP(z) + \gd |\gpwfk(z)|^2
\end{equation}
the ``shifted'' hamiltonian~\cite{bib:kramer.epjd-3.247.2003}.

Each value of the lattice wave number $k$ fixes a solution for
Eq.~\eqref{eq:gpe-k}. To solve it with the corresponding boundary conditions we
express the function $\gpwfk(z)$ in its complex form
\begin{equation}
  \label{eq:wave-function-form}
  \gpwfk(z) = \sqrt{\density(z)} e^{i \phase(z)},
\end{equation}
where the function $S(z)$ represents the phase and $\density(z) \equiv
  |\gpwfk(z)|^2 = |\gpwf_k(z)|^2$ is the particle number density as a function
of $z$.

Substituting~\eqref{eq:wave-function-form} in Eq.~\eqref{eq:gpe-k}, we arrive to
a pair of coupled differential equations for the real and imaginary parts in
terms of $\density(z)$ and $\phase(z)$. The equation for the phase is a first
order differential equation,
\begin{equation}
  \label{eq:gpe-imag}
  \partial_z S(z) = -k + \frac{\alpha}{\density(z)},
\end{equation}
where $\alpha$ is a constant of integration. Equation (\ref{eq:gpe-imag}) is
easily integrated through separation of variables, such that the phase is given
by
\begin{equation}
  \label{eq:gpek-phase}
  \phase(z) = \phase_0 - k z + \int_{0}^{z} \frac{\alpha}{\density(z')} dz',
\end{equation}
with $S_0$ a constant of integration. In order to find the phase we require the
solution for the density $\density(z)$, which comes from the real part of the
GPE~\eqref{eq:gpe-k}:
\begin{align}
  \label{eq:gpe-real}
  \nonumber -\frac{\hbar^2}{2m} & \left(\partial^2_{z} r(z) - r(z){(\partial_z S(z))}^2\right) + \gd r(z)^3 +                                            \\
                                & \quad \quad \left(\frac{\hbar^2 k}{m} \partial_z S(z) + \frac{\hbar^2 k^2}{2m} + \VKP(z) \right) r(z) = \chempot r(z).
\end{align}
Here, $r(z) = \sqrt{\density{(z)}}$. We substitute Eq.~\eqref{eq:gpe-imag} in
Eq.~\eqref{eq:gpe-real}, and after rearranging terms we obtain
\begin{equation}
  \label{eq:gpe-real-alt1}
  -\frac{\partial^2 r(z)}{\partial z^2} + \frac{\alpha^2}{r(z)^3} + \frac{2 m}{\hbar^2} \left(\VKP(z) - \chempot \right) r(z) + \frac{2 m \gd}{\hbar^2} r(z)^3 = 0.
\end{equation}
To proceed further, we take into account that the Kronig-Penney is a piecewise
potential with a constant magnitude in each barrier or well region, i.e.,
$\VKP(z) = V_0$ within the barrier and $\VKP(z) = 0$ inside the well. Let's
focus on the barriers region. We multiply Eq.~\eqref{eq:gpe-real-alt1} by
$\partial_z r(z)$ and integrate the equation, arriving to
\begin{equation}
  \label{eq:gpe-real-alt2}
  -\frac{1}{2} \left(\frac{\partial r(z)}{\partial z}\right)^2 - \frac{\alpha^2}{2 r(z)^2} + \frac{m}{\hbar^2} \left(\vzero - \chempot \right) r(z)^2 + \frac{m \gd}{2 \hbar^2} r(z)^4 - \sigma = 0,
\end{equation}
where $\sigma$ is a second constant of integration. Finally, we multiply this
equation by $-8 r(z)^2$; after some algebraic steps, we obtain the corresponding
differential equation for the density:
\begin{equation}
  \label{eq:gpe-density}
  {\left(\frac{d\density}{dz}\right)}^2 = \frac{4m\gd}{\hbar^2} \density^3 +
  \frac{8m}{\hbar^2}(\vzero - \mu) \density^2 - 8 \sigma \density - 4 \alpha^2.
\end{equation}
An analog procedure for the
wells region gives a similar equation, but with $\vzero = 0$, so the
differential equation~\eqref{eq:gpe-density} changes only in the quadratic term.

The ODE~\eqref{eq:gpe-density} has a set of analytical solutions given by the
Jacobi elliptic functions~\cite{bib:abramowitz-stegun.1964}. The explicit form
of the density is
\begin{equation}
  \label{eq:gp-density}
  \density(z) = \density[\mathrm{off}] + 4 \ellipm \lambda^2 \, \jacobisn^2 \left(
  \sqrt{\frac{4 m \gd}{\hbar^2}} \lambda (z - z_\mathrm{off}) \, \vert \,  \ellipm
  \right),
\end{equation}
where the function $\jacobisn(u \vert \ellipm)$ is the Jacobi elliptic sine. The
factor $\ellipm$ is a real number known as the elliptic modulus,
$\density[\mathrm{off}]$ is a constant offset on the value of $\density(z)$,
while $\lambda$ is a parameter that fixes the amplitude of spatial density
variations. Equation~\eqref{eq:gp-density} defines a whole family of functions
whose properties are deeply linked to the value of $\ellipm$, the value of
$\gd$, and $\lambda$.

\subsection{Boundary conditions}%
\label{sec:boundary-conditions}

\begin{SCfigure}[50][t]
  \includegraphics[width=0.6\linewidth]{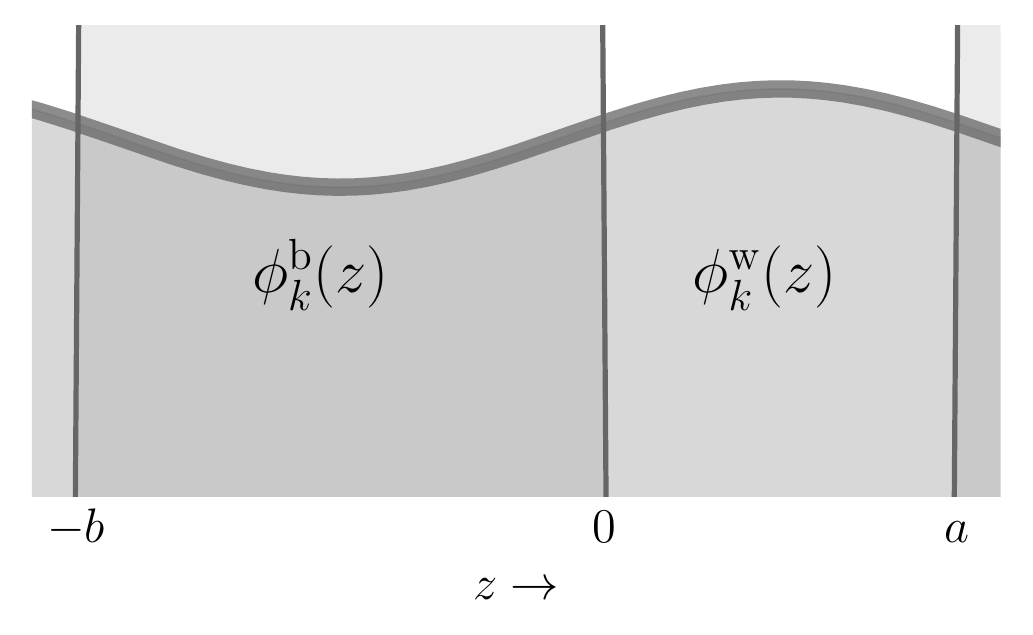}
  \caption{(Color online) Identification of the wave function by region. The
    darker regions correspond to the potential barriers (superscript ``b''),
    while the white regions, to the wells (superscript ``w'').}%
  \label{fig:gpe-wf-parts}
\end{SCfigure}

The solution for the density~\eqref{eq:gp-density} assumes that the potential
magnitude remains constant over the interval of $z$ being evaluated. Hence, we
have two solutions for $\gpwfk(z)$: one within the ``well'' (w) regions and
another within the ``barrier'' (b) regions. Then,
\begin{equation}
  \gpwfk(z) = \begin{cases}
    \gpwfkw(z) = \sqrt{\densityw(z)} e^{i \phasew(z)}, & \VKP(z) = 0   \\
    \gpwfkb(z) = \sqrt{\densityb(z)} e^{i \phaseb(z)}, & \VKP(z) = V_0
  \end{cases}
\end{equation}
We consider also that: both functions must match in a smooth way at the
interface of each potential barrier, the periodic nature of the potential and
that the system is infinite. In Fig.~\ref{fig:gpe-wf-parts} we show the behavior
of the function $\gpwfk(z)$ in both well and barrier regions. Within the
barriers $\gpwfkb(z)$ has a depletion which is complemented by the accretion in
$ \gpwfkw(z)$ within the wells.

A direct consequence of the nature of the potential is that we have a density
function and a set of parameters $\density[\mathrm{off}]$, $\ellipm$, $\lambda$
and $z_\mathrm{off}$, as well as a phase function and parameters $S_0$ and
$\alpha$, for each region. These sets are related by the boundary conditions
imposed on the system.  We can exploit the periodicity of $\gpwfk(z)$ to focus
our analysis to a single period of the system fixed at the origin $z = 0$, which
extends from $z = -b$ to $z = a$. In this picture the edge of the barrier is
located at the origin, therefore the boundary condition for continuity is
$\gpwfkb(0) = \gpwfkw(0)$. This equality results in the conditions
\begin{align}
  \label{eq:dens-cont}
  \densityb(0)          & = \densityw(0),                       \\
  \label{eq:phase-cont}
  \phasew_0 - \phaseb_0 & = 2 n_s \pi, \quad n_s \in \mathbb{Z}
\end{align}
The first of these equations forces the density to be continuous at $z = 0$,
while the second states that the difference of phase between regions is discrete
and equal to an integer multiple of $2 \pi$. The periodicity of $\gpwfk(z)$ can
be stated as $\gpwfkb(-b) = \gpwfkw(a)$, which in turn implies that
\begin{align}
  \label{eq:dens-periodic}
  \densityb(-b) & = \densityw(a)                                                                                                                                     \\
  \label{eq:bec-momentum}
  k (a + b)     & = 2 n_{\mathrm{s}} \pi + \int_{-b}^{0} \frac{\alpha^{\mathrm{b}}}{\densityb(z')} dz' + \int_{0}^{a} \frac{\alpha^{\mathrm{w}}}{\densityw(z')} dz',
\end{align}
with $\alpha^{\mathrm{w}}$ in the wells and $\alpha^{\mathrm{b}}$ in the
barriers. In addition, the derivative of $\gpwfk(z)$ must be continuous at $z =
  0$, i.e., $\partial_z \gpwfkb(0^{-}) = \partial_z \gpwfkw(0^{+})$, which is
equivalent to
\begin{align}
  \label{eq:dens-deriv-cont}
  \partial_z \densityb(0^{-}) & = \partial_z \densityw(0^{+}) \\
  \label{eq:current-cont}
  \alpha^{\mathrm{b}}         & = \alpha^{\mathrm{w}}.
\end{align}
Finally, the derivative of $\gpwfk(z)$ must be also periodic, hence
\begin{equation}
  \label{eq:dens-deriv-periodic}
  \partial_z \densityb(-b) = \partial_z \densityw(a).
\end{equation}
Conditions~\eqref{eq:dens-cont} to~\eqref{eq:dens-deriv-periodic}, along with
the normalization~\eqref{eq:gpwf-norm-period} define the complete set of
solutions for the wave function of the condensate $\gpwfk(z)$.

The definition and properties of the Jacobi elliptic functions permit us to
express the normalization~\eqref{eq:gpwf-norm-period} in closed-form. For this,
we can define the integral
\begin{align}
  \label{eq:bec-norm-func}
  \nonumber N(z) & = \int_{0}^{z} \density(z') \, dz'                                                                                                                                                          \\
                 & = (\density[\mathrm{off}] + 4 \lambda^2) z \; - \frac{4 \lambda}{\sqrt{4 m \gd / \hbar^2}} \Big( \mathcal{E}\left( u(z) | \ellipm \right) - \mathcal{E}\left( u(0) | \ellipm \right) \Big),
\end{align}
as the average number of particles contained in the interval $[0, z]$. The
function $\mathcal{E}(u(z) | \ellipm)$ is the incomplete elliptic integral of
the second kind with argument $u(z) = \sqrt{4 m \gd / \hbar^2} \lambda (z -
  z_{\mathrm{off}})$ expressed in canonical form accordingly
to~\cite{bib:abramowitz-stegun.1964}. Then the normalization condition becomes
\begin{equation}
  \label{eq:density-norm-parts}
  N = \int_{-b}^{0} \densityb(z') \, dz' + \int_{0}^{a} \densityw(z') \, dz',
\end{equation}
where each of the integrals can be evaluated using the
equation~\eqref{eq:bec-norm-func}. Analogously, the integral
in~\eqref{eq:gpek-phase} becomes
\begin{equation}
  \label{eq:bec-momentum-func}
  \int_{0}^{z} \frac{dz'}{\density(z')} = \frac{1}{\sqrt{4 m \gd / \hbar^2} \lambda \density[\mathrm{off}]} \; \times \Big( \Pi (n_\mathrm{j}; u(z) | \ellipm) - \Pi(n_\mathrm{j}; u(0) | \ellipm ) \Big),
\end{equation}
where $\Pi(n_\mathrm{j} ; u(z) | \ellipm)$ is the incomplete elliptic integral
of the third kind of order $n_\mathrm{j} = -4 \ellipm \lambda^2
  /\density[\mathrm{off}]$. Then the quasi-momentum $k$ in~\eqref{eq:bec-momentum}
can be expressed in terms of~\eqref{eq:bec-momentum-func}.

\section{Ground state density profile and chemical potential}%
\label{sec:density-profile}

The periodic structure as well as the interaction between particles have
notorious effects on the density of the condensate, even in the regime where
there is no relative velocity of the gas with respect to the potential frame.
\begin{figure}[b!]
  \begin{minipage}[b]{0.495\textwidth}
    \centering
    \includegraphics[width=0.75\linewidth]{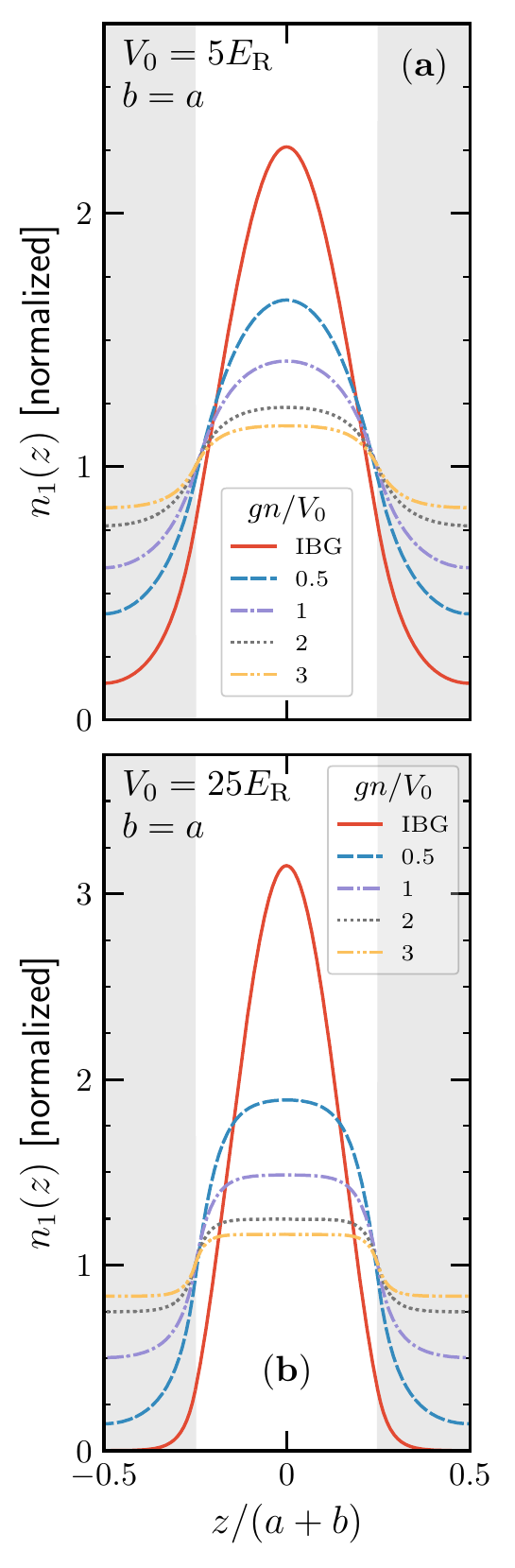}
  \end{minipage}
  \begin{minipage}[b]{0.495\textwidth}
    \centering
    \includegraphics[width=0.75\linewidth]{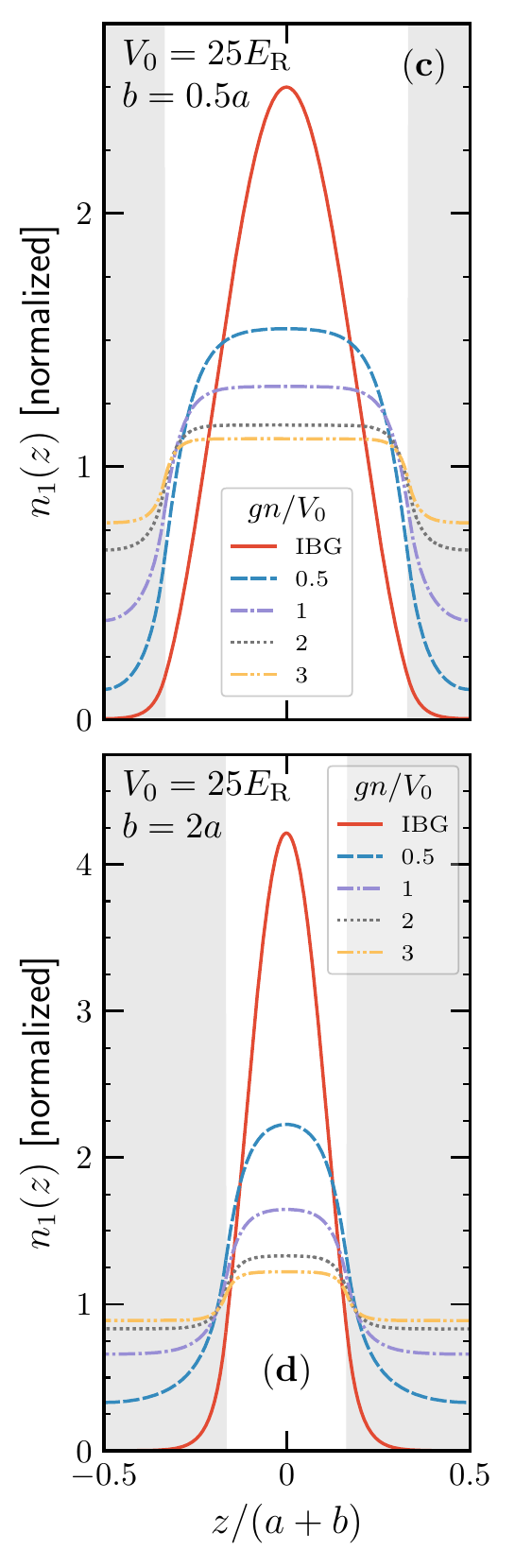}
  \end{minipage}
  \caption{(Color online) Ground state density profile as a function of $z$ for
    for different values of the repulsive interaction strength and for different
    geometries. Top curve (solid) corresponds to the ideal Bose gas. The
    following (dashed) curves from top to bottom correspond to $\gd \avgdensity
      = 0.5, 1, 2, 3$ times $\vzero$. Dark regions indicate the location of the
    potential barriers.}%
  \label{fig:bec-density-r-1}
\end{figure}
In Figs.~\ref{fig:bec-density-r-1}a and~\ref{fig:bec-density-r-1}b we show the
behavior of the density as a function of the position $z$ for two square
lattices with potential barrier height $\vzero = 5\Er$ and $\vzero = 25\Er$,
respectively. The geometric ratio of the potential, defined as $r = b / a$, is
equal to unity since $b = a$. In both plots we have calculated the density
profile for several values of a repulsive two-body interaction strength
$\intfactor$, including the interactionless Bose gas ($\gd\avgdensity = 0$)
affected only by the periodic potential. We show the density profile for
$\intfactor = 0.5, 1, 2, 3$ times $\vzero$. For all cases the density has a
maximum value at the midpoint of the well region, and a minimum at the midpoint
of the barrier region. As the repulsive interaction between particles increases,
we observe that the density variations in a spatial period diminish in such a
way that the average value of the density approaches to unity. This occurs
because the interaction between particles dominate the repulsive effect of the
potential barriers, reducing the particle localization in the wells. An opposite
effect appears when we increase the barrier height, keeping constant the
interaction between particles, then the density profile raises in the well
regions while diminish inside the barriers.
Also, the geometry of the lattice has a significant influence on the density
profile, as it is shown in the Figs.~\ref{fig:bec-density-r-1}c
and~\ref{fig:bec-density-r-1}d, where we have calculated $\density(z)$ for two
different nonsquare lattices with geometric ratios $b = 0.5a$ and $b = 2a$,
respectively, while keeping constant the potential height $\vzero = 25 \Er$. The
results show a greater particle localization in the wells when the potential
barriers becomes wider, since the barrier repulsion dominate over the repulsive
interactions between particles. Eventually, as the repulsive interaction
increases, the density profile becomes flat.

We can obtain a relationship between the energy~\eqref{eq:gpe-energy} and the
chemical potential directly from the {\GP} equation. Multiplying both sides
of~\eqref{eq:gpe-k} by $\gpwfk^{*}(z)$ and integrating over a potential period,
we arrive to
\begin{equation}
  \label{eq:gpe-chem-pot}
  \mu_k = \frac{1}{N} \int_{0}^{l} \gpwfk^{*}(z) \gpeham^{k} \gpwfk(z) \, dz.
\end{equation}
We recognize that the only difference between the total energy and chemical
potential arises from the nonlinear term. It follows that $\gpenergy[k]$ and
$\mu_k$ are related by
\begin{equation}
  \label{eq:gpe-energy-chem-pot-rel}
  \mu_k = \frac{\gpenergy[k][\gpwfk]}{N} + \frac{\gd \avgdensity}{2} \int_{0}^{1}
  {\Big( \densitydl[1k](\tilde{z}) \Big)}^2 \, d\tilde{z},
\end{equation}
where $\densitydl[1k](\tilde{z}) = \left| \gpwfk(\tilde{z})\right|^2 / n$, with
$\tilde{z} = z / (a + b)$. The second term of the r.h.s.\
of~\eqref{eq:gpe-energy-chem-pot-rel} accounts for the two-body interactions in
the gas. It vanishes for $\gd = 0$, therefore, $\mu_k = \gpenergy[k]/N$, which
is the exact result for the energy per particle of a noninteracting Bose gas at
zero temperature. In this case, the chemical potential can be obtained using the
dispersion relation of an ideal Bose gas subject to a Kronig-Penney
potential~\cite{bib:rodriguez.jltp-3.144.2016}.

\begin{figure}[b]
  \begin{minipage}[b]{0.495\linewidth}
    \centering
    \includegraphics[width=0.8\linewidth]{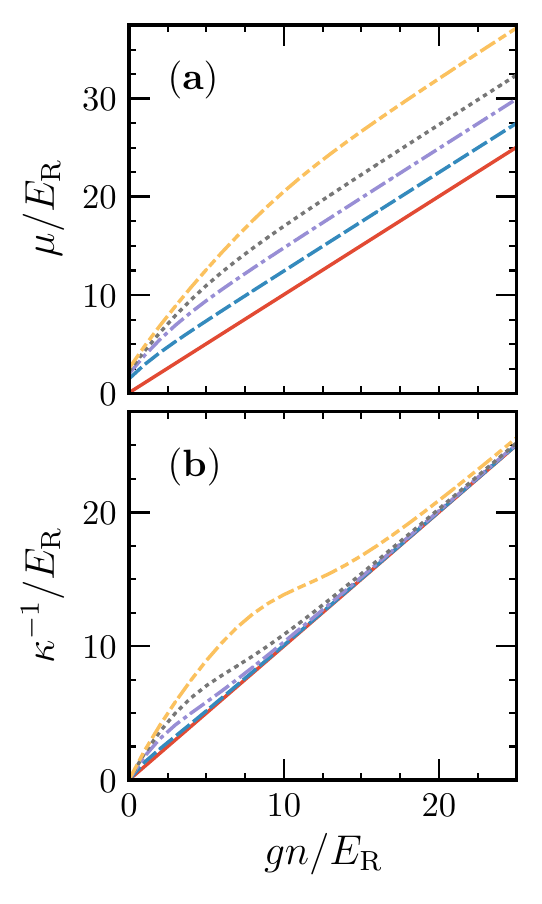}
  \end{minipage}
  \begin{minipage}[b]{0.495\linewidth}
    \centering
    \includegraphics[width=0.8\linewidth]{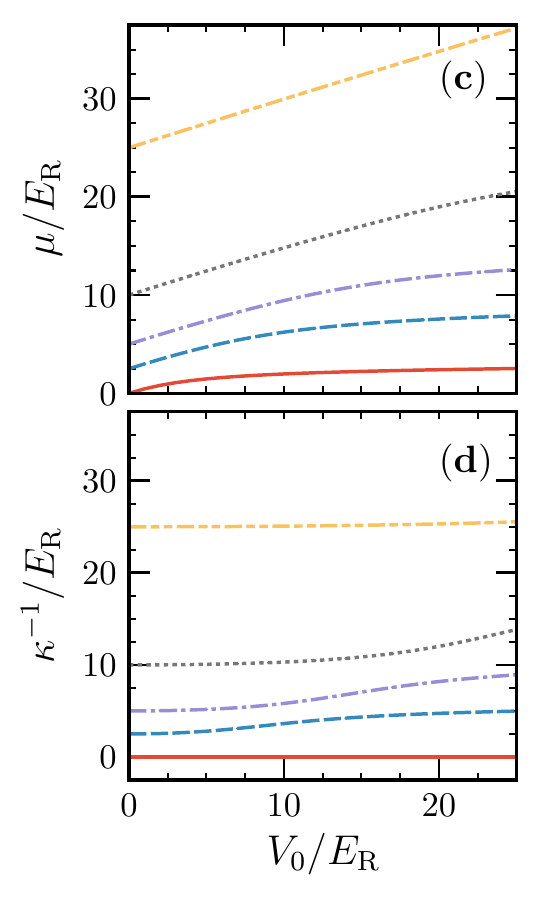}
  \end{minipage}
  \caption{ (Color online) Chemical potential and inverse of the compressibility
    of the ground state $k = 0$ for a square lattice $b = a$. \textbf{(a)} and
    \textbf{(b)}: as a function of the interaction strength. The solid line
    corresponds to the free gas. Dashed lines, from top to bottom, correspond to
    lattice heights $\vzero = 25, 15, 10$ and $5$ times $\Er$, respectively.
    \textbf{(c)} and \textbf{(d)}: as a function of the lattice height. The
    solid line corresponds to the noninteracting gas. Dashed lines, from top to
    bottom, correspond to $\gd \avgdensity = 25, 10, 5$ and $2.5$ times $\Er$,
    respectively.}%
  \label{fig:mu-kz_kappa-gs_r-1}
\end{figure}

We calculated the chemical potential and the compressibility for the state $k=0$
which are shown in Fig.~\ref{fig:mu-kz_kappa-gs_r-1}. We observe that the
chemical potential is a monotonic, increasing function of the interaction
parameter $\gd \avgdensity$ (Fig.~\ref{fig:mu-kz_kappa-gs_r-1}a). In the limit
when the interaction goes to zero, $\chempot$ tends correctly to the value of
the ideal Bose gas, which in general is nonzero due to the presence of the
lattice: the repulsive effect of the barriers raises the chemical potential,
since we require more energy to add a single particle to the system. Obviously,
larger potentials raise $\mu$ even further. When the interaction is strong
enough so that the kinetic energy becomes small compared to the potential
energy, we can obtain a closed-form formula for $\chempot$. Neglecting the
kinetic energy term in Eq.~\eqref{eq:gpe-hamiltonian-k} and using the
normalization condition~\eqref{eq:gpwf-norm-period}, it follows that
\begin{equation}
  \label{eq:chemp-pot-thomas-fermi}
  \mu_{\mathrm{TF}} = \gd \avgdensity + \frac{r}{1 + r} \vzero,
\end{equation}
which is the chemical potential in the so-called Thomas-Fermi limit. It
basically implies that, when the interaction strength is strong enough, the
chemical potential increases as the corresponding one of the free, but
interacting, Bose gas, plus a shift due to the external potential. When the
potential $\vzero$ vanishes, Eq.~\eqref{eq:chemp-pot-thomas-fermi} reduces to
the case of the free, interacting Bose gas. However,
Eq.~\eqref{eq:chemp-pot-thomas-fermi} is not valid anymore when $\vzero$
approximates to $\intfactor$, since the spatial variations of $\gpwfk(z)$ grow
and the kinetic term (proportional to $|\partial_z \gpwfk(z)|$) of the {\GP}
equation becomes significant. We can see this change in the
Fig.~\ref{fig:mu-kz_kappa-gs_r-1}c, where the linear dependence of $\chempot$ on
$\vzero$ is lost as $\intfactor$ becomes smaller.

The compressibility of the gas $\kappa$ is related to the chemical potential by
the relation
\begin{equation}
  \label{eq:compressibility}
  \kappa^{-1} = \avgdensity \frac{\partial \mu}{\partial \avgdensity},
\end{equation}
where we implicitly assume that $\mu$ is calculated for a specific Bloch state
with a fixed momentum $\hbar k$. For a free, homogeneous interacting Bose gas,
the chemical potential is $\mu = \intfactor$~\cite{bib:pitaevskii.oup.2016}, so
the inverse of the compressibility is $\kappa^{-1} = \gd \avgdensity$. In
general, the inverse compressibility will grow at the same rate, i.e.,
$\kappa^{-1} \approx \intfactor$, in both $\intfactor \to 0$ and $\intfactor \gg
  \vzero$ limits. In the latter case, the compressibility resembles the one of the
free gas because the relatively large interactions screen out the effects of the
external potential (Fig.~\ref{fig:mu-kz_kappa-gs_r-1}b). For a nonzero lattice
height in an intermediate range of $\intfactor$ the compressibility will deviate
from the free Bose gas behavior, since the presence of the external potential
reduces the compressibility of the gas due to the repulsive nature of the
barriers (Fig.~\ref{fig:mu-kz_kappa-gs_r-1}d).

\section{Nonlinear energy band structure}%
\label{sec:energy-spectrum}

\begin{SCfigure}[50][b]
  \includegraphics[width=0.4\linewidth]{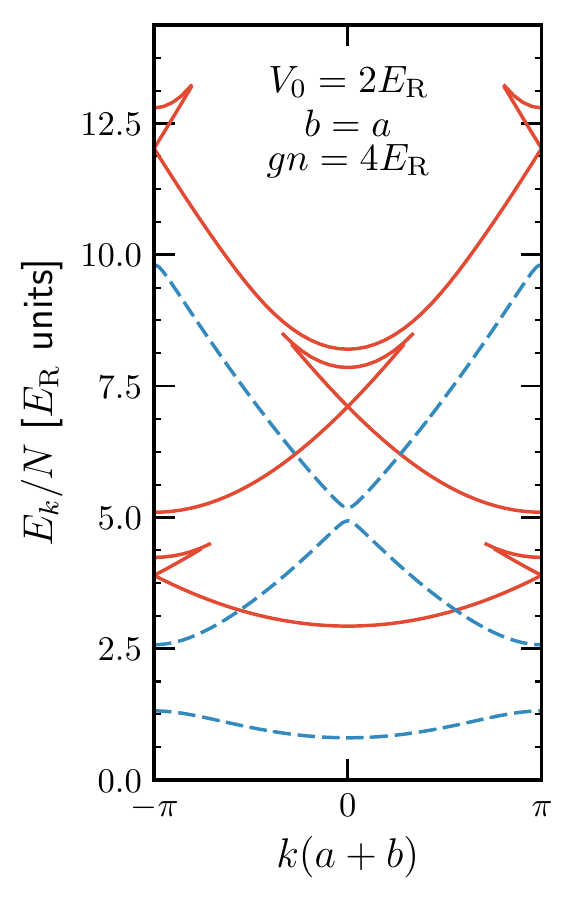}
  \caption{(Color online) First three bands of the energy spectrum for a square
    lattice in the first Brillouin zone. The solid lines correspond to the
    interacting system, the dashed lines show the non-interacting spectrum. The
    swallow tails in the band structure become larger as the interaction factor
    between particles to potential ratio increases.}%
  \label{fig:energy-spectrum-square}
\end{SCfigure}

The presence of the nonlinear term in the {\GP} equation~\eqref{eq:gpe-k} alters
the band spectrum of the non-interacting case in striking ways. The most notable
phenomena is the appearance of energy loops the so-called ``swallow
tails''~\cite{bib:machholm_pra-67.053613.2003}.
Figure~\ref{fig:energy-spectrum-square} shows the nonlinear band spectrum for a
square lattice $b = a$ with potential height $\vzero = 2 \Er$. In addition, the
lattice contain a strongly repulsive condensate where $\gd \avgdensity = 4 \Er$.
The characteristic shape of swallow tails is readily visible.
The energy loops belong to a specific band, and appear in a regular way: at the
end of the first Brillouin zone for odd bands, or at the center for even bands.\
They become larger as the repulsive interaction magnitude increases respect to
the lattice height. The swallow tails emerge because, as the interaction
increases, two states appear that share the same crystal momentum but different
energies. Both of these states $\gpwfk(z)$ are minimizers of the energy
functional~\eqref{eq:gpe-energy} subject to the normalization
condition~\eqref{eq:gpwf-norm} for a fixed chemical
potential~\cite{bib:rogel-ejp.2013}. The origin of the swallow tails is a
consequence of the change of the the energy landscape of the Bose gas, i.e., the
shape of $\gpenergy[][\gpwf]$ as a function of $\gpwf(z)$. The nonlinear term in
the {\GP} results in a situation where the system has more than one state that
minimize the energy
functional~\eqref{eq:gpe-energy}~\cite{bib:mueller-pra.6.063603.2002}. The
appearance of two local minima implies (only by physical considerations) the
existence of a local maximum in the landscape between both minima. The state
with maximum energy will lie in the upper portion of the swallow tail (for a
fixed momentum $k$), while the remaining state will lie in the lower portion as
it has less energy.
\begin{figure}[b]
  \centering
  \begin{minipage}[b]{0.495\linewidth}
    \centering
    \includegraphics[width=0.8\linewidth]{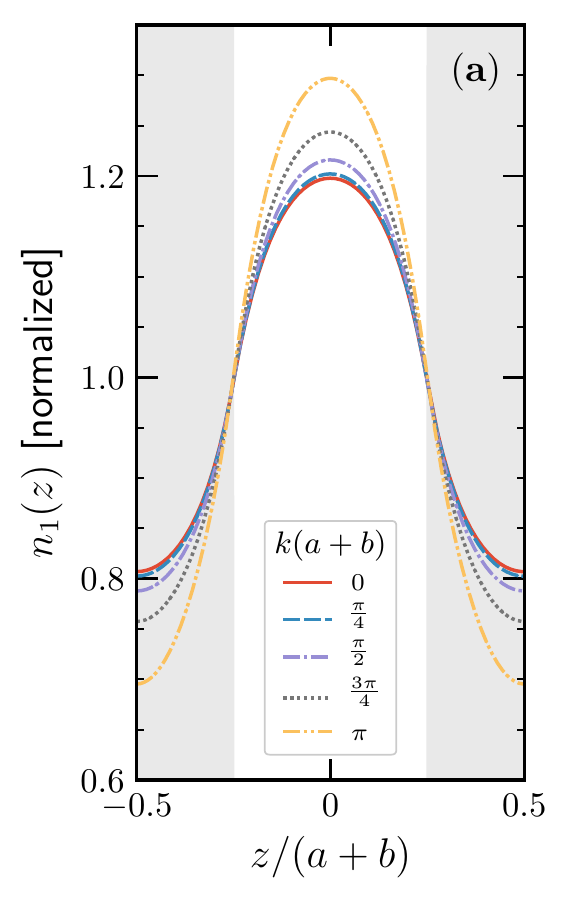}
  \end{minipage}
  \begin{minipage}[b]{0.495\linewidth}
    \centering
    \includegraphics[width=0.8\linewidth]{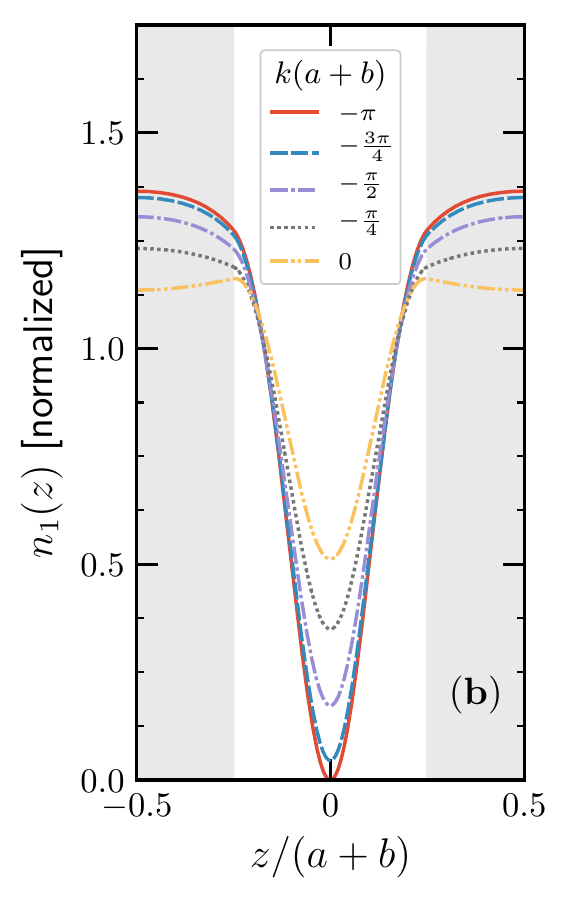}
  \end{minipage}
  \caption{(Color online) Density profile as a function of $z$ for some excited
    states in \textbf{(a)}: first energy band and \textbf{(b)}: second energy
    band, of the system $\vzero = 2\Er$, $b = a$ and $gn = 4 \Er$. Each curve
    corresponds to a different momentum $k$.}%
  \label{fig:bec-density-excited-states-square}
\end{figure}
The appearance of the swallow tails depends mainly on the ratio $\vzero / \gd
  \avgdensity$, and, at a lesser extent, on the geometric ratio $b / a$.  For
optical lattices the swallow tail for the first band appears when the
interaction factor $\intfactor$ becomes equal (or greater) than the lattice
height~\cite{bib:machholm_pra-67.053613.2003}. For upper bands there is not a
similar analytical relation.

Figure~\ref{fig:bec-density-excited-states-square}a and
~\ref{fig:bec-density-excited-states-square}b show the density profiles of some
states in the first an second energy bands (within the first Brillouin zone),
respectively, for a square lattice ($b = a$) with height $\vzero = 2\Er$ and
interaction strength $\gd \avgdensity = 4\Er$. The energy spectrum for this
system is the one shown in Fig.~\ref{fig:energy-spectrum-square}. The states
plotted in Fig.~\ref{fig:bec-density-excited-states-square}a lie in the first
band, in the lower part of it. The solid line represents the ground state $k=0$,
while the rest of the curves have increasing values of the momentum, up to $k (a
  + b) = \pi$. As we can see, the maximum of the density at the origin decreases
as the momentum increases, while the density reduces in the midpoint of the
potential barriers. This corresponds to a greater kinetic energy as the density
spatial variations are magnified.
The curves in the Fig.~\ref{fig:bec-density-excited-states-square}b show the
density variations for a set of states that lie in the lower part of the second
energy band. In this case the variations of the density get reduced as the
momentum grows, from an initial state at the left edge of the Brillouin zone,
with $k(a + b) = -\pi$. This state has the peculiarity that it becomes zero at
the origin, in the midpoint between two potential barriers. The periodicity of
the Bloch states implies that there are zero density surfaces at $z = j (a +
  b)$, with $j$ an integer. This states form an array of so-called ``dark
solitons''~\cite{bib:machholm_pra-67.053613.2003,bib:tsuzuki_jltp.4.441.1971}.

\section{Subwavelength optical lattices as experimental {\KPacr} potential
  realizations}%
\label{sec:sub-wavelength-opt-latt-kp}

In recent years, the experimental realization of optical lattices with
subwavelength spatial structure~\cite{bib:wang_prl-120.083601.2018} has open a
way to study the physics of quantum manybody fluids subject to periodic
potentials that closely resemble the well known {\KP} potential in the Dirac
$\delta$-function limit~\cite{bib:merzbacher_qm.jws.2nd.1969}. In this
approximation, the width of the potential barriers $b$ goes to zero, and the
potential magnitude $\vzero$ goes to infinity, but the product $\vzero b$
remains constant. Then, the {\KPacr} potential becomes a succession of
$\delta$-functions centered in the positions $j l$, being $j$ an integer and
$\KPperiod$ the {\KPacr} potential period, as well as the separation between two
contiguous deltas. The expression~\eqref{eq:kp-potential-def} for $\VKP(z)$
becomes the Dirac-comb potential,
\begin{equation}
  \label{eq:dirac-comb-potential}
  V_{\mathrm{DC}}(z) = \vzero b \sum_{j=-\infty}^{\infty} \delta(z - j \KPperiod),
\end{equation}
where the finite, constant value $\vzero b$ is the area below a single barrier
of the {\KPacr} potential. In the context of the Dirac comb potential, it can be
interpreted as a measure of the impermeability (the strength) of a single Dirac
delta barrier. When $\vzero$ becomes zero we recover the homogeneous, free
interacting Bose gas. On the opposite side, as the delta strength becomes larger
the system resembles more a succession of independent wells of infinite walls,
each one having a width $\KPperiod$.
Since $\vzero b$ is an energy times a length, we can redefine the delta strength
as $\vzero b = s \Er \KPperiod$. Although this definition seems somewhat
arbitrary, it is very useful when we write Eq.~\eqref{eq:dirac-comb-potential}
in terms of the dimensionless length $z' = z / \KPperiod$,
\begin{equation}
  \label{eq:dirac-comb-scaled}
  V_{\mathrm{DC}}(z) = s \Er \sum_{j=-\infty}^{\infty} \delta(z' - j),
\end{equation}
where we used the scaling property of the delta function, $\delta(\KPperiod z) =
  \delta(z) / \KPperiod$. Then, in Eq.~\eqref{eq:dirac-comb-scaled} the factor
$s$ represents the dimensionless strength of the potential in $\Er$ units, and
relates the parameters of our multi-rods model with the Dirac-comb potential
parameters. As an ubiquitous potential, the stationary states of mean-field
BEC subject to an external, Dirac-comb potential~\eqref{eq:dirac-comb-scaled}
have been previously
studied~\cite{bib:theodorakis_jpa-13.4835.1997,bib:weidong-li_phys-rev-e-70.016605.2004,bib:seaman-pra-3.033609.2005,bib:dong_lp-2.190.2007}.
\begin{figure}[t]
  \begin{minipage}[b]{0.495\linewidth}
    \centering
    \includegraphics[width=0.83\linewidth]{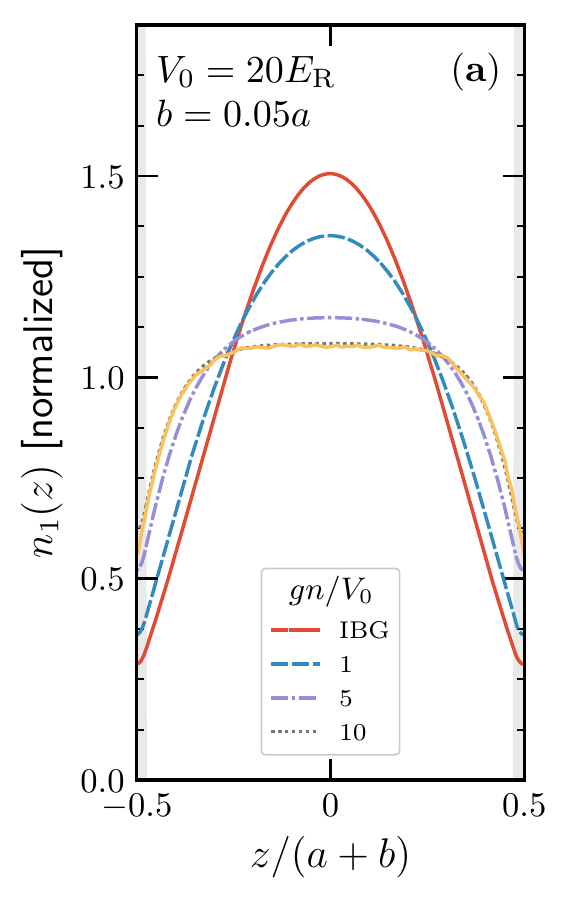}
  \end{minipage}
  \begin{minipage}[b]{0.495\linewidth}
    \centering
    \includegraphics[width=0.8\linewidth]{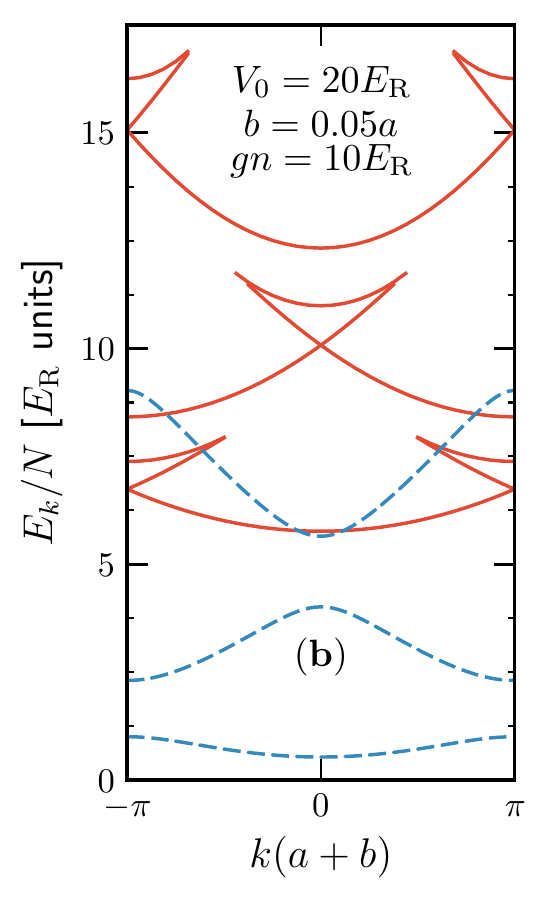}
  \end{minipage}
  \caption{(Color online) \textbf{(a)}: Ground state density as a function of
    $z$ as the {\KP} potential approaches to the Dirac-comb potential with
    $\vzero b = \Er a$. Top curve (solid) corresponds to the ideal gas. The
    following (dashed) curves from top to bottom correspond to $g \avgdensity =
      1, 5, 10$ times $\Er$. Results from~\cite{bib:seaman-pra-3.033622.2005} for
    $g \avgdensity = 10 \Er$ (solid, yellow line). \textbf{(b)}: Energy band
    structure in the first Brillouin zone, for lattice parameters close to the
    limit of the Dirac-comb potential. The solid lines correspond to the
    interacting system, the dashed lines show the noninteracting spectrum.}%
  \label{fig:bec-density-energy-delta}
\end{figure}

In Fig.~\ref{fig:bec-density-energy-delta}a we present the density profile of
the Bose gas as the {\KP} potential approaches to the Dirac-comb potential.  We
can see that the shape of the density profile becomes flatter in the middle of
the lattice cell as the interaction strength increases. There is a sharp change
around the edged of the cell, where the Dirac deltas are located. Results are
very similar for both systems, even when they have very different values of
$\vzero$ and $b$.  Our results confirm that, for potentials equal or greater
than $\vzero = 20 \Er$, and for ratios of the order or smaller than $b / a =
  0.05$, the multi-rods potential is a very close representation of the Dirac-comb
potential with strength $s = 1$. The calculated density profiles are in very
good agreement with those in Fig. 7 of~\cite{bib:seaman-pra-3.033622.2005}.
Naturally, for a stronger Dirac comb potential, i.e., for a greater value of
$s$, the established thresholds for $\vzero$ and $b$, such that a potential
barrier represents a delta barrier, will change. In
Fig.~\ref{fig:bec-density-energy-delta}b we show the energy spectrum for a
relatively large interaction parameter $\intfactor = 10 \Er$. The swallow tails
are notorious, and appear in all the plotted bands. The difference with the
spectrum of the ideal gas is complete.

Based on the previous analysis of the Dirac comb potential, we attemp to model
the optical potential~\cite{bib:wang_prl-120.083601.2018}
\begin{equation}
  \label{eq:sub-wav-opt-pot}
  V_{\mathrm{OL}}(z) = \frac{\epsilon^2 \cos^2(\kopt z)}{{(\epsilon^2 + \sin^2(\kopt z))}^2} \Er
\end{equation}
in the limit when $\epsilon \ll 1$. Under this condition, $V_{\mathrm{OL}}(z)$
becomes a lattice of narrow barriers spaced by (a period) $\lambdaopt / 2$, with
a peak value of $\Er / \epsilon^2$ and width at half maximum scaling of $\Delta
  = \epsilon \lambdaopt / 2 \pi \ll
  \lambdaopt$~\cite{bib:lacki_prl-117.233001.2016}. For $\epsilon \ll 1$, this
potential has a subwavelength spatial structure that is a very close
approximation of the Dirac comb potential~\eqref{eq:dirac-comb-potential} with
strength $s = 1 / (2 \epsilon)$.
We realize our analysis by fixing the {\KPacr} barrier width as $b = 0.05a$, and
equating the {\KPacr} potential period with the optical lattice period, i.e.,
$\KPperiod = \lambdaopt / 2$. Then, to get the corresponding barrier height
$\vzero$, we equate the barrier area $\vzero b$ with the area of
$V_\mathrm{OL}(z)$ over a potential period for $\epsilon = 0.14$. This procedure
results in a barrier height of $\vzero = 74.3 \Er$, somewhat larger than the
peak of $V_\mathrm{OL}(z)$, which is $51.02 \Er$.

First, we have calculated the predicted nonlinear band structure of the
condensate for the system with $\intfactor = 10 \Er$, whose results are plotted
in Fig.~\ref{fig:energy-spectrum-dirac-comb}. This is a relatively strong
condensate, with an energy spectrum that significantly deviates from the ideal
one. Swallow tails are not significant in the spectrum, however.

\subsection{Ground state energy of the Lieb-Liniger Bose gas within a subwavelength lattice}

\begin{SCfigure}[50][t]
  \includegraphics[width=0.4\textwidth]{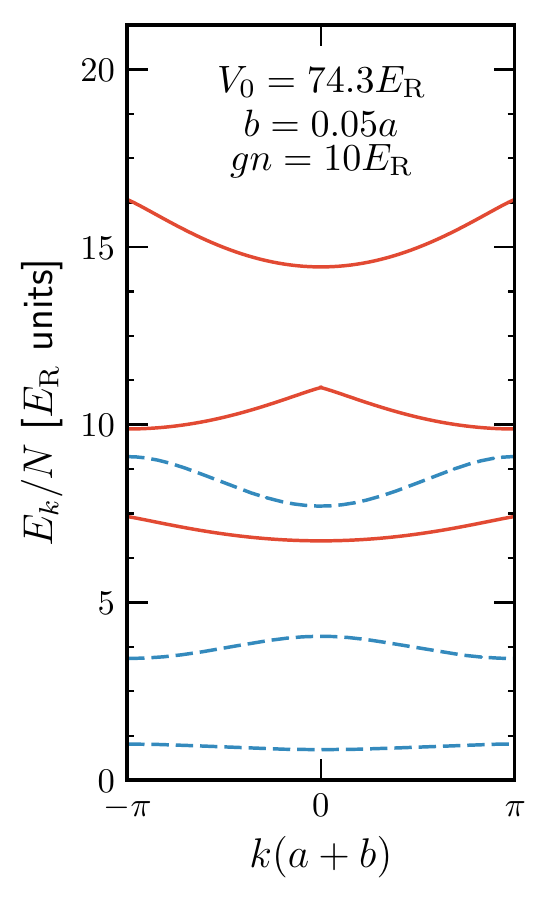}
  \caption{(Color online) Band structure in the first Brillouin zone. {\KPacr}
  potential parameters are chosen so that the system closely resembles the
  optical potential~\eqref{eq:sub-wav-opt-pot} for $\epsilon = 0.14$. The solid
  lines correspond to the interacting system, the dashed lines show the
  noninteracting spectrum.}%
  \label{fig:energy-spectrum-dirac-comb}
\end{SCfigure}

\begin{SCfigure}[50][b]
  \includegraphics[width=0.5\textwidth]{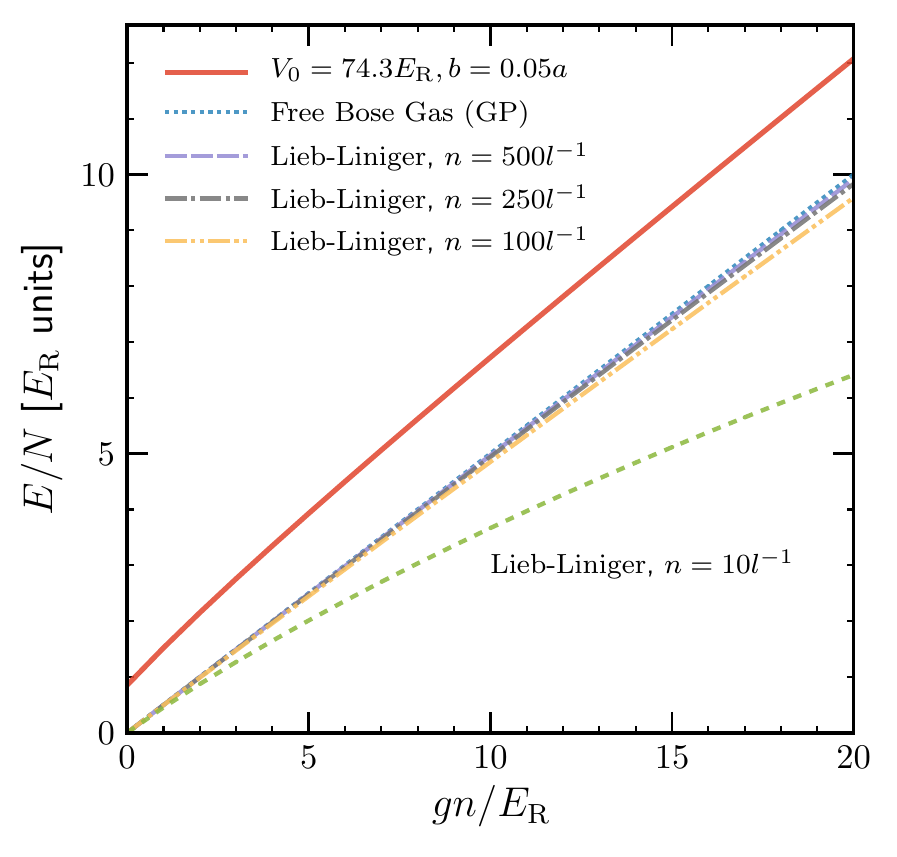}
  \caption{Energy of the ground state as a function of the interaction strength.
    The solid line corresponds to the Bose gas subject to the multirods
    potential close to the Dirac comb limit.}%
  \label{fig:energy-comp-dirac-comb-ll}
\end{SCfigure}

The extent to which a 1D Bose gas with interactions of the order of $gn = 20
  E_R$ or less, is well modeled by the GPE will depend not only on the value of
$g$, but also on the average linear density of the gas $\avgdensity$. Unlike
the 3D Bose gas, the weakly interacting regime for a 1D Bose gas corresponds
to a high average density of the gas, while low densities correspond to the
strongly interacting
regime~\cite{bib:pitaevskii.oup.2016,bib:dunjko_prl.86.5413.2001}.  The exact
description of a 1D Bose gas with contact-like, repulsive interactions is
given by Lieb-Liniger (LL) model~\cite{bib:lieb_phys-rev-130.1605.1963} and
the LL parameter $\gamma = m \gd / \hbar^2 \avgdensity$, which must satisfy
$\gamma \ll 1$ in order to the GPE picture to be valid. We have $\gamma = {(nl
  / \pi)}^{-2} / 2 \times \intfactor / \Er$.
Figure~\ref{fig:energy-comp-dirac-comb-ll} shows the ground state energy of
the gas predicted by the LL theory as a function of $\intfactor$, for average
densities of $\avgdensity = 500$, $250$, $100$ and $10$ times
$\KPperiod^{-1}$. We compare these results with the energy per particle
$\intfactor / 2$ predicted by the {\GPEacr} for the homogeneous Bose gas.
Results show that only in the high density case $\avgdensity = 500 l^{-1}$ the
  {\GPEacr} gives accurate results over the full interval of interaction
$\intfactor$. Higher densities should provide results even more accurate. For
$\avgdensity = 100 l^{-1}$ we can see small discrepancies between the
  {\GPEacr} predictions and the LL theory.  Numerical results indicate that for
$\avgdensity = 250 l^{-1}$ the {\GPEacr} and LL exact results are almost
identical, and it can be taken as a lower limit for suitable densities;
greater values of $\avgdensity$ are within the range of typical experimental
densities for Bose gases in 1D
regime~\cite{bib:mewes_prl.77.416.1996,bib:kruger_prl.105.265302.2010}. In the
presence of the subwavelength optical potential the ground state energy of the
Bose gas is shown in Fig.~\ref{fig:energy-comp-dirac-comb-ll} by the solid,
red line. The lattice raises the energy with respect to the free Bose gas, and
its dependence is not linear with $\intfactor$. Then, the observed energy band
structure of the gas for $\intfactor = 10 \Er$ should be similar to the one
shown in Fig.~\ref{fig:energy-spectrum-dirac-comb}.

\section{Conclusions}%
\label{sec:conclusions}

We have studied a 1D interacting Bose gas at zero temperature subject to a
periodic, multi-rods potential by quasi-analytically solving the
Gross-Pitaevskii equation. We model the periodic structure using a Kronig-Penney
potential, which has the remarkable property of having analytical solutions for
the wave function of the condensate. In this work, we focused on the Bloch state
type solutions of the GP equation.

We were able to find analytical expressions for the wave function of the
condensate. The density profile, the normalization condition, and the complex
phase can be expressed in terms of the Jacobi elliptic functions, as well as in
terms of the incomplete elliptic integrals of the first, second and third kind.
We have obtained the density profile, the chemical potential and the energy
spectrum of an interacting trapped Bose gas. The energy spectrum consists of
bands separated by prohibited regions, like the spectrum of the trapped ideal
gas. However, we found that the nonlinear spectrum may strongly differ from the
ideal one since the first one shows loops, or ``swallow tails'', at the edges of
the first Brillouin zone for odd bands and at the center for even bands.  No
analytical expression for a threshold of the appearance of swallowtails was
obtained so that further research in this subject is required. We obtained the
chemical potential as well as the compressibility, numerically, as functions of
the potential magnitude $\vzero$ and the interaction parameter $\intfactor$.
When the interaction between particles is relatively large compared to the
lattice height, our results agree satisfactorily with those predicted by the
Thomas-Fermi approximation, where we have obtained closed-form expressions for
the chemical potential and the compressibility.

Our periodic lattice becomes the Dirac comb potential when the potential
barriers become very high and very thin, but the area below them remains finite.
Although we have not used a mathematical strict expression for a Dirac
$\delta$-potential, we found that for barriers as thin as $b = 0.05a$ the
  {\KPacr} potential is a very good approximation of the Dirac comb potential. In
this case, we can reproduce the results for the density profile and the energy
spectrum done in previous studies for the Bose gas in Dirac comb potentials.
Moreover, because we have full control of the width and height of the potential
barriers, we have employed our periodic ultranarrow rods model to predict the
energy band structure of an interacting 1D BEC in optical lattices with
subwavelength spatial structure. By comparison of the ground state energies of
the free Bose gas calculated exactly (LL case) and approximately (GP case), for
the trapped Bose gas, we give a lower estimate for the average linear density of
$250$ bosons per potential period, such that the {\GP} equation results should
be considered accurate in the full interaction interval shown in
Fig.~\ref{fig:energy-comp-dirac-comb-ll}.

We acknowledge partial support from grants PAPIIT-DGAPA-UNAM IN-107616 and
IN-110319, and CONACyT 221030.

\bibliographystyle{spphys}
\bibliography{main}

\end{document}